\documentclass[11pt,a4paper]{article}
\usepackage[dvips]{graphicx}
\usepackage{subfigure}
\input epsf
\begin{document}

\title{Incommensurate structures studied by a modified Density Matrix 
Renormalization Group Method}
\author{A.~Gendiar and A.~\v{S}urda\\
Institute of Physics, Slovak Academy of Sciences, D\'{u}bravsk\'{a} cesta 9,\\
SK-842 28 Bratislava, Slovak Republic}

\maketitle{}

\begin{abstract}
A modified density matrix renormalization group {\it DMRG} method is introduced and 
applied to classical two-dimensional models: the anisotropic triangular nearest-neighbor 
Ising (ATNNI) model and the anisotropic triangular next-nearest-neighbor Ising (ANNNI) model.
Phase diagrams of both models have complex structures and exhibit incommensurate 
phases. It was found that the incommensurate phase completely separates the disordered
phase from one of the commensurate phases, i. e. the non-existence of the Lifshitz point
in phase diagrams of both models was confirmed.
\end{abstract}

{\bf PACS:} 05.70.Fh, 64.70.Rh, 02.60.Dc, 02.70.-c

\renewcommand{\theequation}{\arabic{equation}}
\setcounter{equation}{0}
\section{Introduction}

\qquad
In 1992 S. R. White \cite{Whi} invented the density matrix renormalization
group ({\it DMRG}) technique in real space which has been mostly used for  
diagonalization of one-dimensional (1D) quantum chain spin Hamiltonians. 
Three years later T. Nishino \cite{Nis} applied this numerical technique to 
classical spin 2D models. The {\it DMRG} method is based on  renormalization 
of the transfer matrix. It is a variational method that 
maximizes the partition function using a limited number of degrees of freedom 
and the variational state is written as a product of local matrices \cite{Var}.

A similar method (cluster transfer matrix method) for classical spin models was developed
by one of us, where the variational state is written as a product of local functions \cite{Sur,Su3}.
In this paper we shall compare the results of both methods.

The {\it DMRG} technique proceeds in two steps. In the first 
one the infinite system method ({\it ISM}) pushes both ends of the system
from each other 
and enlarges the system by two sites at each iteration. In the second one the finite system 
method ({\it FSM}), in which the system has fixed size, improves the values of calculated
physical quantities by several left and right moves (sweeps)  yielding very accurate 
results \cite{Whi}. 

{\it DMRG} has been used for many various quantum models. It provides 
results with remarkable accuracy for larger systems than it is possible
to study using standard diagonalization methods.
The 2D classical systems treated by the {\it DMRG} method exceeds the classical Monte 
Carlo approach in accuracy, speed, and size of the systems \cite{Kar}. 
In spite of this, too few works have been still done for 2D classical spin models 
by the {\it DMRG} technique \cite{Nis,Drw,Ni2}. 
A further {\it DMRG} improvement of the classical systems is based 
on Baxter's corner transfer matrix \cite{CTM}, the {\it CTMRG} 
\cite{CTMRG}, and its generalization to any dimension \cite{3DN}.

The aim of this paper is to investigate two classical models that exhibit 
incommensurate phases, namely the anisotropic next-nearest-neighbor Ising 
(ANNNI) model \cite{Sur} and the anisotropic triangular nearest-neighbor
Ising (ATNNI) model \cite{Dom}. The incommensurate
phases were studied by many various theoretical approaches. The free fermion 
approximation revealed IC phase in 2D classical ANNNI model \cite{Vil} 
(also \cite{Dom} in ATNNI model), 2D incommensurate crystals \cite{Pok}. 
Incommensurate structures has been discussed in various topics:
2D C-IC phase transition \cite{Sch}, 2D quantum ANNNI model \cite{Bar}, 
ANNNI model in $d>2$ dimensions \cite{Fis}, and by analyzing  1D sine-Gordon 
model \cite{Hal} where the authors found no Lifshitz point. 

A modified {\it DMRG} method which can be applicable to more complicated systems, 
namely the ANNNI and ATNNI models will be developed. Both models 
are characterized by non-symmetric transfer matrices. The way 
of how to use {\it DMRG} in such a case will be described in this paper. 
Since most work done in {\it DMRG} has been devoted to the models
described by symmetric transfer matrices (2D classical models) or
hermitian quantum Hamiltonians (1D quantum models), we show a modified {\it DMRG}
for treating non-symmetric transfer matrix in the ATNNI model.                                                            
In particular, the existence or non-existence of the Lifshitz point in the ATNNI 
model will be studied and its phase diagrams will be constructed.

Paper is organized as follows: in Sec. 2 we compare two various methods in using of 
the {\it DMRG} technique; in Sec. 3 we describe the ATNNI and ANNNI models;
Sec. 4 contains the modified {\it DMRG} algorithm for the ATNNI model;
in Sec. 5 we present our results and in Sec. 6 the results will be summarized and discussed.

\section{The DMRG technique for 2D spin systems}

\qquad
For special values of interaction constants both ATNNI and ANNNI models
can be reduced to the Ising model. In this case we can compare our approximate
{\it DMRG} calculations with the exact results for infinite 2D models. 

Exact results can be relatively easily obtained  for 1D models, e.g. strips
of finite width. They provide   a good opportunity for testing our
methods, as well.

The {\it DMRG} technique as a numerical real-space method, is in fact always applied to finite
systems. However, in dependence on the size of the system, it can yield 
approximate descriptions of 1D or 2D infinite systems. 

In case of relatively narrow strips the {\it DMRG} calculations are consistent
with the exact calculations: they yield zero-order parameter and reproduces
well  the
largest and the second largest eigenvalues of the transfer matrix of the
system. Comparison of the exact and approximate values  for Ising and ATNNI
models on a semi-infinite strip of width $L=16$ with periodic boundary
conditions \cite{And} for various approximations are given in Tables \ref{tab1} and \ref{tab2}. 
\begin{table}[tb]
\caption {\it The largest eigenvalues $\lambda_1$ and the second largest eigenvalues $\lambda_2$
of the transfer matrices calculated with the {\it DMRG} technique for the 
Ising model with periodic boundary conditions. The 
dimension of the transfer matrix $N$ depends on the size of the block-spin variable 
\cite{Nis}.
The last line of the table contains the eigenvalues of the transfer matrix
obtained by the exact diagonalization method.}
\vspace{0.5cm}
\centerline{
\begin{tabular}{|c|c|c|c|c|}
\hline
\multicolumn{5}{|c|}{\bf Ising model with periodic boundary conditions}\\
\hline
 & \multicolumn{2}{|c|}{Ordered phase} &
\multicolumn{2}{|c|}{Disordered phase} \\
N & \multicolumn{2}{|c|}{$T=2.1$} &
\multicolumn{2}{|c|}{$T=2.4$} \\
\cline{2-5}
& $\lambda_1$ & $\lambda_2$ & $\lambda_1$ & $\lambda_2$ \\
\hline
 400 & 7.00331679 E+06 & 6.9180364 E+06 & 1.76379494 E+06 & 1.5477034 E+06 \\
 1600 & 7.03990343 E+06 & 6.9742292 E+06 & 1.76702461 E+06 & 1.5769844 E+06 \\
 3600 & 7.03991836 E+06 & 6.9742595 E+06 & 1.76704324 E+06 & 1.5771406 E+06 \\
 6400 & 7.04001144 E+06 & 6.9743133 E+06 & 1.76710592 E+06 & 1.5771736 E+06 \\
 10000 & 7.04001146 E+06 & 6.9743135 E+06 & 1.76710593 E+06 & 1.5771740 E+06 \\
\hline
\hline
65536 & 7.04001165 E+06 & 6.9743146 E+06 & 1.76710598 E+06 & 1.5771799 E+06 \\
\hline
\end{tabular}
\label{tab1}
}
\end{table}
\begin{table}[tb]
\caption {\it The two largest eigenvalues $\lambda_1$ and $\lambda_2$ of the transfer matrix for 
the ATNNI model are calculated with the {\it DMRG} technique for periodic boundary 
conditions \cite{And}. Data obtained by the exact diagonalization method are shown 
in the last line.}
\vspace{0.5cm}
\centerline{
\begin{tabular}{|c|c|c|c|c|}
\hline
\multicolumn{5}{|c|}{\bf ATNNI model with periodic boundary conditions}\\
\hline
& \multicolumn{2}{|c|}{Commensurate phase} &
\multicolumn{2}{|c|}{Disordered phase} \\
N & \multicolumn{2}{|c|}{$T=0.9$, $H=2.0$} &
\multicolumn{2}{|c|}{$T=1.2$, $H=2.0$} \\
\cline{2-5}
& $\lambda_1$ & $\lambda_2$ & $\lambda_1$ & $\lambda_2$ \\
\hline
 400 & 3.8724 E+13 & 2.4898 E+12 & 5.8940 E+10 & 1.0267 E+10 \\
 1600 & 4.0818 E+13 & 3.7378 E+13 & 6.9274 E+10 & 4.5860 E+10 \\
 3600 & 4.0503 E+13 & 3.7013 E+13 & 6.9235 E+10 & 4.5697 E+10 \\
 6400 & 4.0560 E+13 & 3.7033 E+13 & 6.9330 E+10 & 4.5120 E+10 \\
 10000 & 4.0556 E+13 & 3.6996 E+13 & 6.9328 E+10 & 4.4892 E+10 \\
\hline
\hline
65536 & 4.0530 E+13 & 3.6884 E+13 & 6.9312 E+10 & 4.4368 E+10 \\
\hline
\end{tabular}
}
\label{tab2}
\end{table}
We see that the first two eigenvalues  of the superblock  transfer matrices in
the {\it DMRG} method  are very close to the exact values  despite the sizes of superblock
matrices $(N\times N)$ are much less than the size of the exact T-matrix
$(65536\times 65536)$. The calculations for ATNNI model were performed at a
moderate magnetic field $H=2$. At higher magnetic fields we frequently encountered
problems with complex conjugated pairs of two largest transfer matrix eigenvalues
in the {\it DMRG} calculations. Consequent symmetrizing of the density matrix 
(presented in \cite{Car}) did not improved calculations.

A 1D model at non-zero temperature does not display any phase transitions.
Nevertheless, the value of the critical temperature for the corresponding 2D
model can be found from finite-size scaling (FSS) considerations \cite{Nig}. This
approach represents first of {\it two methods} we shall use for determination of
the critical temperature. Its value for Ising model, derived from comparing two
rescaled semi-infinite strips of width $L=12$ and 14 with periodic boundary
conditions, are the following: 

i) $T_{\rm c}=2.26987$ from the exact eigenvalues of the transfer matrices
of the size $N=4096$ and $N=16384$,

ii) $T_{\rm c}=2.27008$ from the eigenvalues of the {\it DMRG} superblock transfer
matrices of the size $N=1024$

\noindent comparing with the exact critical temperature of 2D Ising model 
$T_{\rm c}=2.26918\dots$ \cite{Bax}. 

The second method for determination of the critical temperature is provided 
by {\it DMRG} calculations on 2D systems, large in both directions. Here, below the
critical temperature a spontaneous symmetry breaking occurs, i.e. the order
parameter acquires non-zero values and tends to zero at the critical point.
The {\it DMRG} method behaves in a mean-field-like manner. As now $T_c$ can 
be determined directly, no finite size scaling is necessary.     
Its accuracy improves with the size of the superblock transfer matrix
as follows: 

i) $T_c=2.275$ for $N=1024$,

ii) $T_c=2.272$ for $N=3600$,

iii) $T_c=2.2692$ for $N=19600$.

For lower orders of approximation the accuracy of this method is worse than of
the first one, but it converges faster to the exact value.
For the ATNNI and ANNNI models in the phases with broken symmetry, the method
explicitly gives the structure of commensurate as well as incommensurate
phase.  In contrast to the FSS method, it is applicable also in the high
magnetic field region and we were able to investigate nearly the whole phase
diagram of the model.

\section{ATNNI and ANNNI models}
{\bf (I) The ATNNI model}

\begin{figure}[ht]
\centering
\subfigure[\hfill]{\scalebox{0.5}{\includegraphics{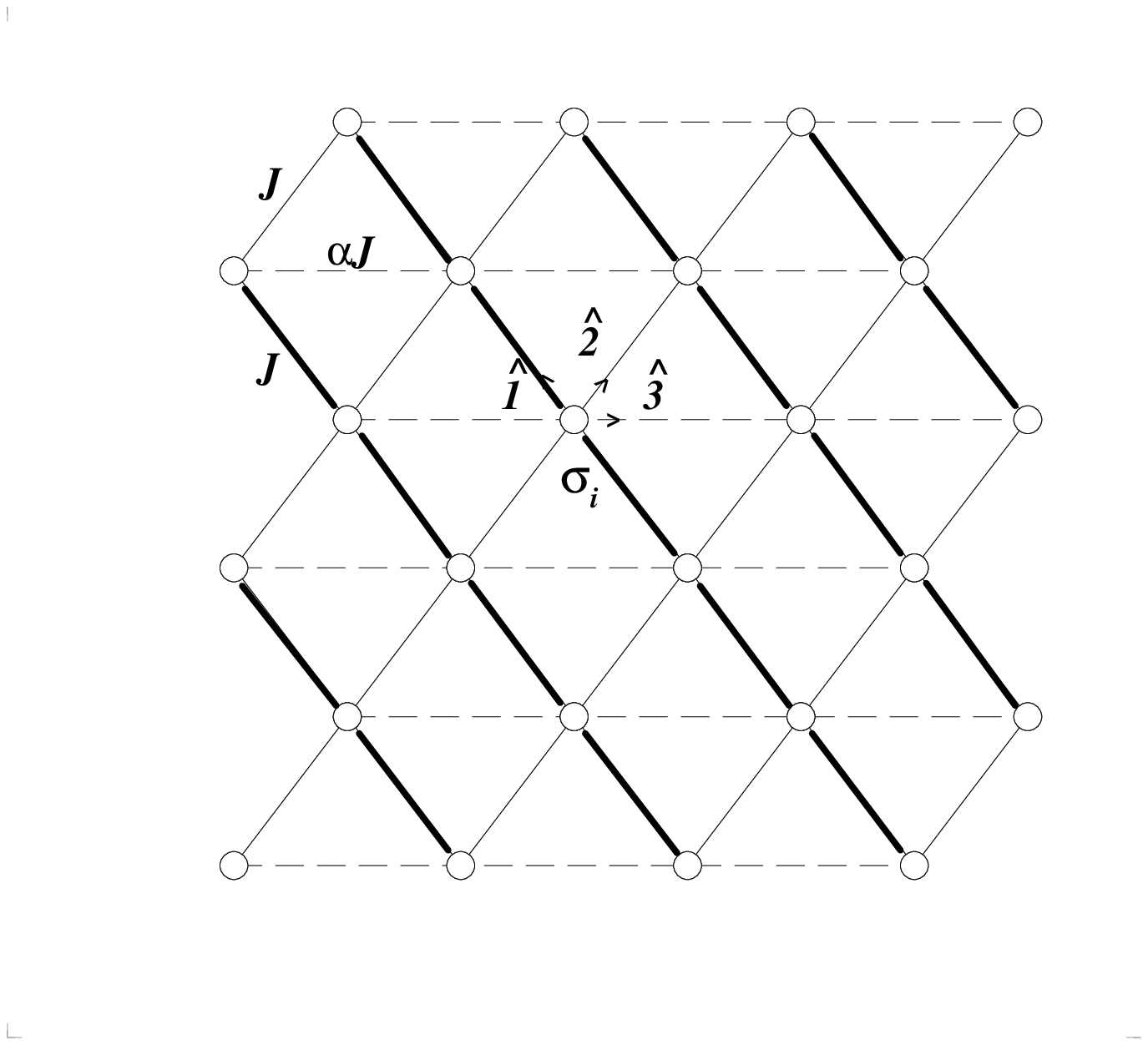}}}
\hfill
\subfigure[\hfill]{\scalebox{0.5}{\includegraphics{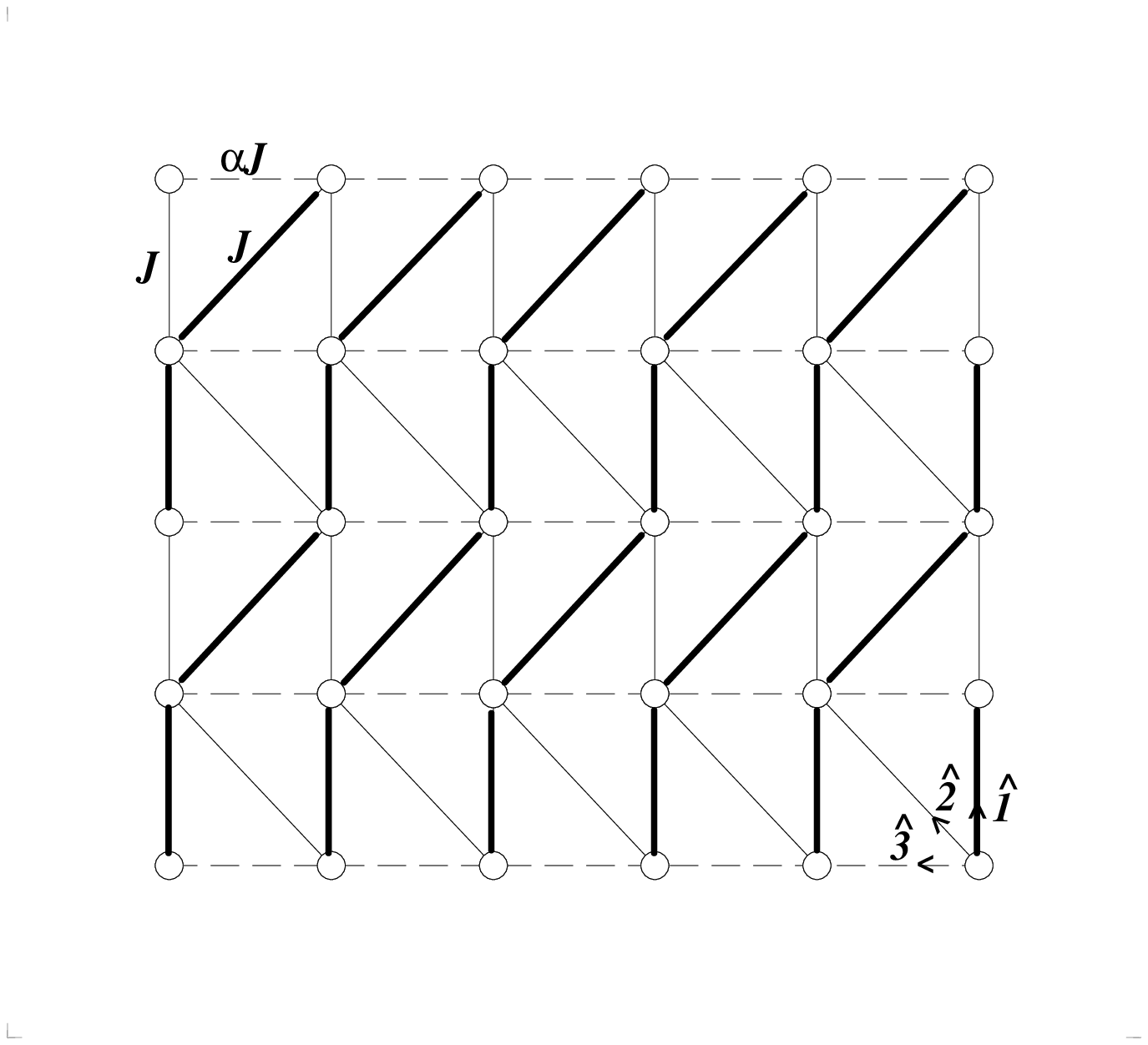}}}
\caption{\it (a) Triangular lattice of the ATNNI model. Along $\hat 3$ direction
(dashed lines) the incommensurate structure appears; (b) The same lattice model 
designed to transfer matrices that are located here as the horizontal strips. Now 
the direction $\hat{1}$ in Figure (a) corresponds to thick lines in Figure (b).}
\label{atnnimodel}
\end{figure}

We consider the two-dimensional classical Ising model with antiferromagnetic interactions
between nearest neighbors on a triangular lattice (the ATNNI model). Its Hamiltonian 
is as follows:
\begin{equation}
{\cal H}=\sum\limits_{i}J\left(\sum\limits_{\hat\delta=\hat 1,\hat 2} \sigma_i 
\sigma_{i+\hat\delta}+\alpha \sigma_i \sigma_{i+\hat 3}\right)-H\sum\limits_i \sigma_i
\end{equation}
with $J>0$, $0<\alpha<1$, and the directions $\hat 1$, $\hat 2$, $\hat 3$ where
$ \sigma_i=\pm 1$, as depicted in Figure \ref{atnnimodel} (a). 

\begin{figure}[ht]
\centerline{\scalebox{0.5}{\includegraphics{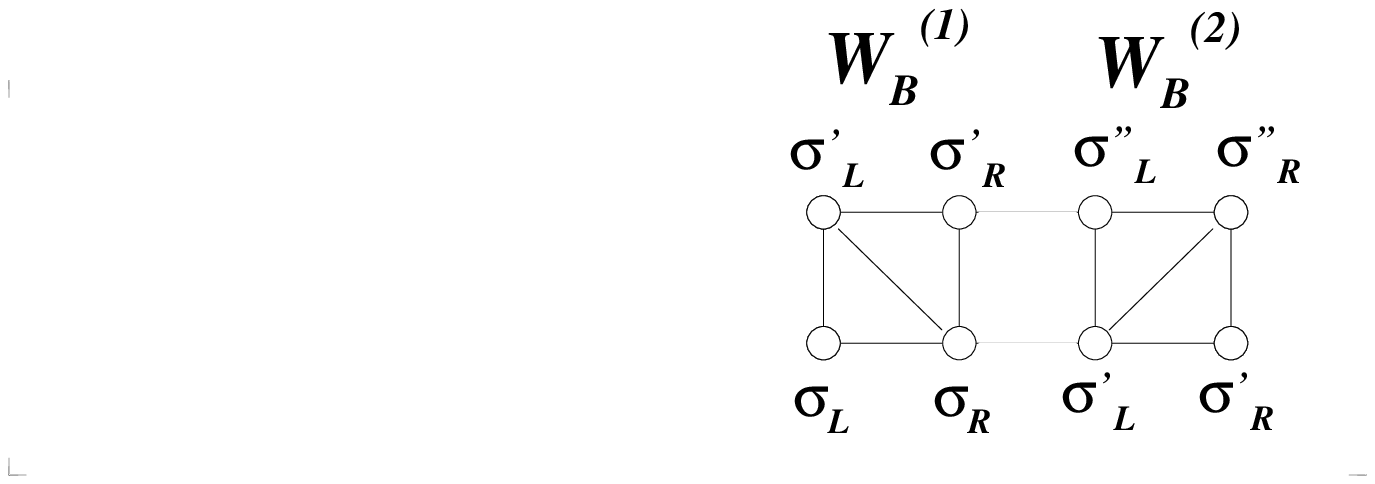}}}
\caption{\it Two different Boltzmann weights that differ from each other by
the orientation of diagonal interactions.}
\label{boltzmannatnni}
\end{figure}

The partition function ${\cal Z}=\sum_{\{\sigma\}}e^{-\beta{\cal H}}$ where 
$\beta={(k_BT)}^{-1}$ can be written as a product of two types
of Boltzmann weights. Each Boltzmann weight $W_B(\sigma_1\sigma_2|
\sigma_1^{\prime}\sigma_2^{\prime})$ is composed of four spins which
interact among themselves as  seen
in Figure~\ref{boltzmannatnni}.

The model is exactly solvable for the external
magnetic field $H=0$. At nonzero temperature ($T_c\approx 1.55$ for 
$\alpha=0.4$ \cite{Dom}), it exhibits the second order phase transition. 
Throughout this paper all numerical calculations are performed
at the fixed constant $\alpha=0.4$, normalization $J=1$, the dimensionless 
temperature $T/J$, and dimensionless ratio $H/T$ in order to compare our 
results with those obtained in \cite{Dom}.

The numerical calculations are based on a diagonalization of two transfer 
matrices (in next Sec. we offer more detailed description of their construction). For this 
purpose we used a square lattice depicted in Figure~\ref{atnnimodel}(b)                                                
which is related to the initial triangular lattice of the ATNNI model as seen in 
Figure~\ref{atnnimodel}(a). We identify direction $\hat{3}$ with the interaction $\alpha J$ 
(Figure~\ref{atnnimodel}). In this direction the incommensurate modulation 
should appear. We use the row-to-row transfer matrices \cite {Dom}.\\
{\bf (II) The ANNNI model}

The ANNNI model is defined on the 2D triangular lattice with  nearest-neighbor
ferromagnetic interactions $J_1<0$ for all three directions and a next-nearest-neighbor
antiferromagnetic interaction $J_2>0$ in one of three directions only. Its Hamiltonian
can be written as
\begin{equation}
{\cal H}=\sum\limits_{i}\left(\sum\limits_{\hat{\delta}=\hat{1},\hat{2},\hat{3}}J_1\sigma_i
\sigma_{i+\hat{\delta}}+J_2\sigma_i\sigma_{i+\hat{4}}\right).
\end{equation}
\begin{figure}[ht]
\centering
\centerline{\scalebox{0.5}{\includegraphics{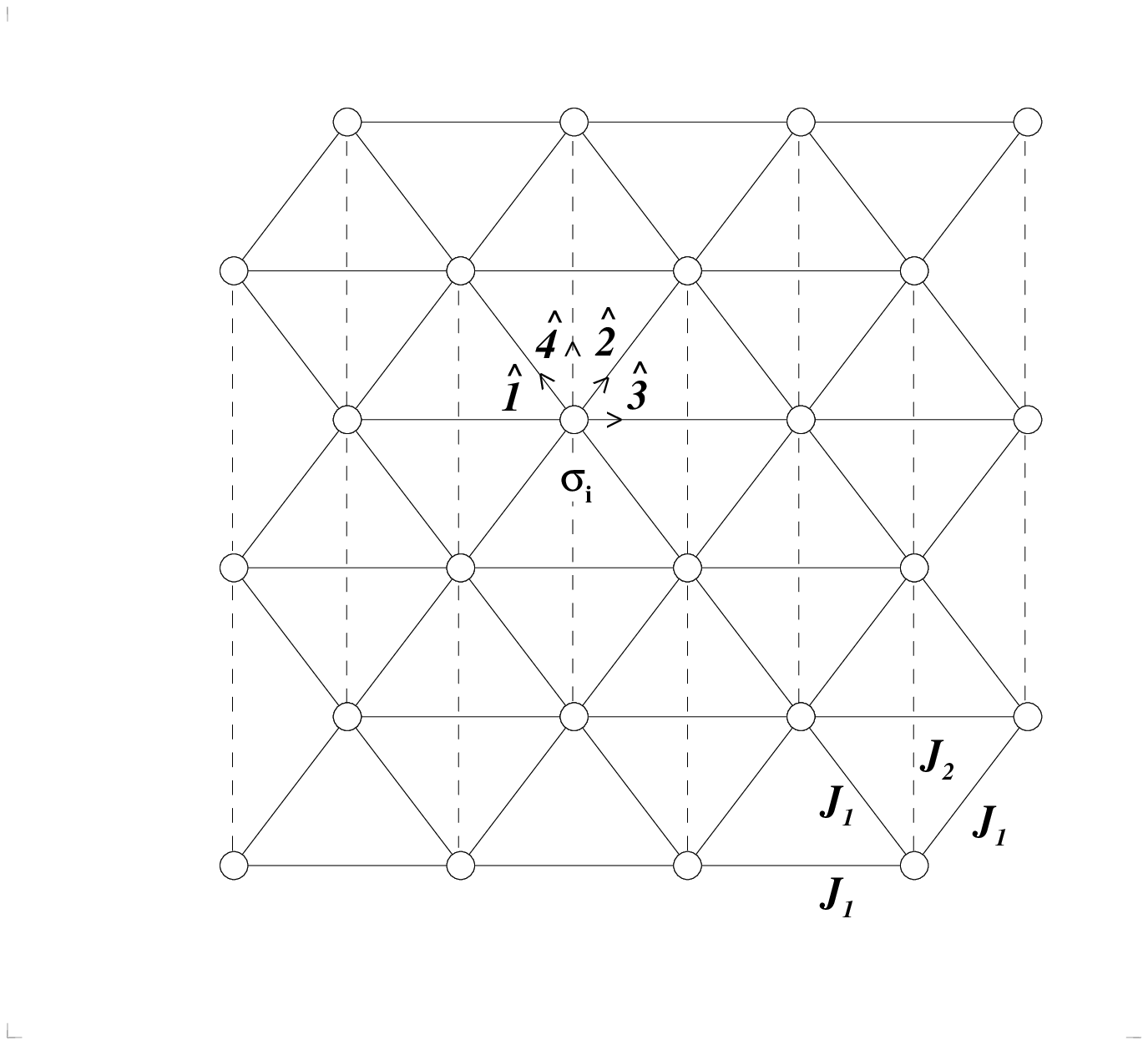}}}
\caption{\it The 2D ANNNI model on the triangular lattice. Directions $\hat{1}$, $\hat{2}$, and $\hat{3}$
characterize the ferromagnetic interaction $J_1$. The next-nearest-neighbor antiferromagnetic
interaction $J_2$ (dashed line) acts in the direction $\hat{4}$ in which the incommensurate
phase appears.}
\label{annnimodel}
\end{figure}

The ANNNI model is usually defined on the square lattice where the 
next-nearest-neighbor interactions, in fact, are equal to zero and the
third-nearest-neighbor ones are non-zero and  antiferromagnetic. The ANNNI model 
on the triangular
lattice is the real Anisotropic Next-Nearest-Neighbor Interaction model with 
non-zero next-nearest-neighbor interactions
and vanishing the third-nearest-neighbor ones.
A frustration of the ANNNI model appears due to the competing interactions. The
ANNNI model was mostly studied on the square lattice \cite{Sel} but it was 
shown by \cite{Sur}
that the properties of the ANNNI model on the triangular lattices remain essentially 
unchanged.

Boltzmann weight is composed of six spins $W_B(\sigma_1\sigma_2\sigma_3|\sigma_1^{\prime}
\sigma_2^{\prime}\sigma_3^{\prime})$.

\begin{figure}[ht]
\centerline{\scalebox{0.2}{\includegraphics{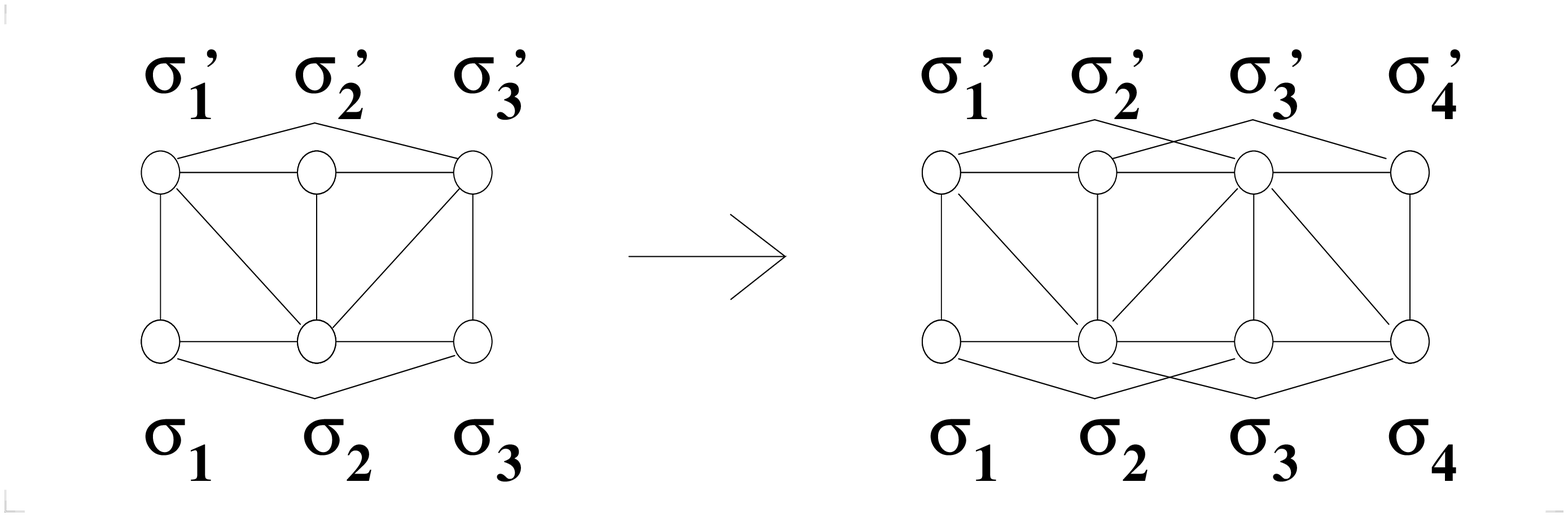}}}
\caption{\it The six-spin Boltzmann weights of the ANNNI model differing from each other by 
the orientation of diagonal interactions. In the DMRG calculation the Boltzmann 
weight defined
on the eight-spin cluster was used. It is composed of two overlaping six-spin Boltzmann
weights.}
\label{boltzmannannni}
\end{figure}

The phase diagram of the ANNNI model (will be discussed in Sec. 5)
consists of four regions: ferromagnetic phase with non-zero total magnetization,
commensurate phase $\langle 2\rangle$ with periodically
alternating spin signs ($\cdots++--++--\cdots$), paramagnetic phase,
and incommensurate phase located between the commensurate and paramagnetic phases.

\section{Modification of the DMRG algorithm}

The {\it DMRG} algorithm for quantum models introduced by White \cite{Whi} was modified
and applied for 2D classical lattice models by Nishino \cite{Nis}. As the ATNNI
and ANNNI model on the triangular lattice lead to non-symmetric transfer
matrices and incommensurate phases, the Nishino's approach has to be
modified further. We shall pursue the second approach discussed in Section 2 --
the {\it DMRG} method applied to very wide strips where the spontaneous symmetry 
breaking occurs. 

The {\it DMRG} method replaces the exact row-to-row transfer matrix of a strip, which is a
product of plaquette Boltzmann weights, by a set of much
smaller superblock transfer matrices for every plaquette.  
The superblock transfer matrix consists of the Boltzmann weight $^iW_B$ of the
plaquette $i$ multiplied by left and right transfer matrices (blocks) $^iT_L$, 
$^iT_R$ which replace all the remaining plaquette Boltzmann weights of the exact transfer 
matrix
to the left and right from the plaquette. The left and right transfer matrices 
are indexed by
left and right spins $\sigma_{L,R}=\pm 1$ of the plaquette, respectively, 
and by block-spin variables $\xi=1,\dots,m$. The number of spin components $m$ 
determines the order of the approximation. For a modulated phase, $^iT_L$ and $^iT_R$ 
differ for 
each plaquette and, in the {\it FSM}, are calculated self-consistently from the transfer 
matrices 
corresponding to neighboring plaquettes. The left block $^{i+1}T_L$  is obtained from 
$^iT_L\,^iW_B$ after a reduction of its matrix size to the original value in a proper way.
The right block $^{i-1}T_R$  is similarly calculated from $^iW_B\,^iT_R$. A calculation of 
the left and right transfer matrices is performed iteratively during a number of
sweeps across the strip.
         
\begin{figure}[ht]
\centerline{\scalebox{0.9}[0.7]{\includegraphics{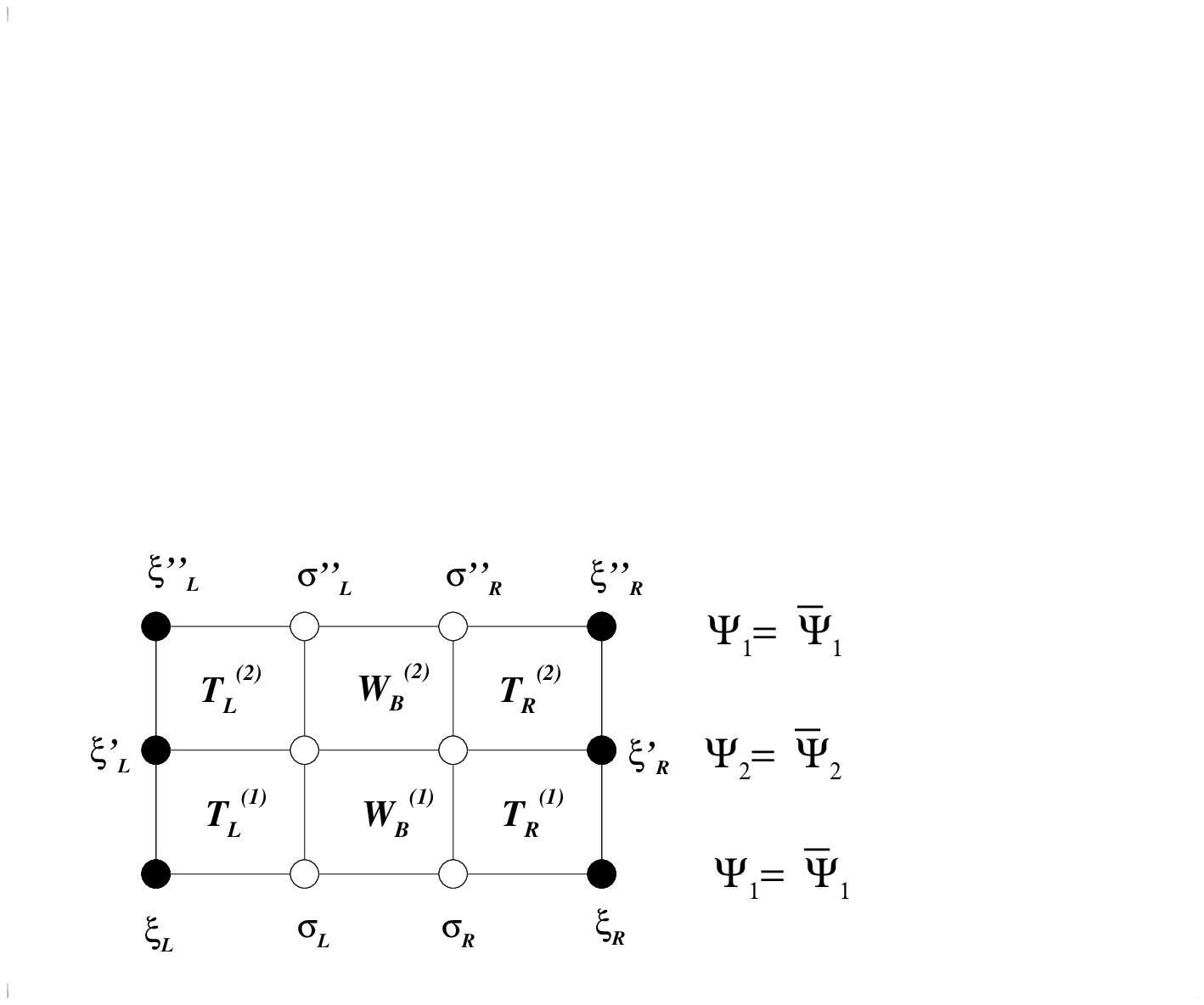}}}
\caption{\it $\Psi_1$, $\overline\Psi_1$ are the right and left eigenvectors that correspond to
the largest eigenvalue of the superblock transfer matrices 
${}^iT^{(1)}=\,^iT_L^{(1)}\,^iW_B^{(1)}\,^iT_R^{(1)}$ and ${}^iT^{(2)}=\,^iT_L^{(2)}\,^iW_B^{(2)}\,^iT_R^{(2)}$,
respectively. Vectors $\Psi_2$ and $\overline\Psi_2$ are needed to calculate a density matrix 
in each {\it DMRG}
iteration.}
\label{atnnitransfermatrix}
\end{figure}

The reduction of the size of the transfer matrices is based on density
matrices that are constructed from the left and right (or in Figure~\ref{atnnimodel}(b) and 
\ref{atnnitransfermatrix} rather upper and
lower) eigenvectors of the superblock matrix \cite{Whi}.

The above-described procedure for homogeneous phases and symmetric transfer
matrices is thoroughly explained in \cite{Nis}. For the ATNNI and ANNNI models the
method should be slightly modified because the transfer matrices are not
symmetric and the structure is modulated in both direction, in one of them,
incommensurately. It is convenient to choose the strip perpendicularly to
the direction of the incommensurate modulation with commensurate structure 
along the strip,
i.e. the strip is orientated in the vertical direction of the lattice shown in
Figure~\ref{atnnimodel}{b}. It is seen that there are two different row-to-row non-symmetric
transfer matrices in the strip. One transfer matrix is constructed from
the Boltzmann weight $W_B^{(1)}$ and the other one from the $W_B^{(2)}$ 
(Figure~\ref{atnnimodel}(b) and \ref{atnnitransfermatrix}).
 
The number of superblock transfer matrices should be equal to at least two, 
as  number of the Boltzmann weights they are attached to. For {\it DMRG} with
spontaneous symmetry breaking and spatially modulated structures, they may be
different for each plaquette of the lattice. The density matrices are not
constructed from the eigenvectors of the superblock transfer matrices but
rather from functions obtained by an iterative procedure 
$\Psi_j=\prod_{k=1}^j T^{(k)}\Psi_1$ starting from $\Psi_1$ given by
suitable boundary conditions. For a homogeneous structure and large $j$ the
function $\Psi_j$ is identical to the eigenvector of the superblock $T$. 

Fortunately, all the commensurate structures of ATNNI model are as a maximum  
of period 2 in the direction of the strip. Then, we have only two superblock transfer
matrices $T^{(1)}=T^{(1)}_L W^{(1)}_B T^{(1)}_R$ and  
$T^{(2)}=T^{(2)}_L W^{(2)}_B T^{(2)}_R$ shown in Fig.~\ref{atnnitransfermatrix}, however,  
they are dependent on the position of the plaquette in the horizontal (perpendicular to 
the strip) direction. Similarly, as a result of the iteration procedure we obtain only
two different functions $\Psi_j$ where $\Psi_1$ is the eigenvector of the matrix 
$T^{(2)}T^{(1)}$
for $j$ even and $\Psi_2$ is the eigenvector of the matrix product $T^{(1)}T^{(2)}$ 
for $j$ odd. 
These both combined matrices are already symmetric and their right  and left
eigenvectors,  $\Psi$ and $\overline\Psi$, respectively, are identical.

Writing down the spin variables explicitly, the right eigenvectors are given by the
equation
\begin{equation}
\sum\limits_{\xi_L\sigma_L\sigma_R\xi_R}T(\xi_L^{\prime\prime}\sigma_L^
{\prime\prime}\sigma_R^{\prime\prime}\xi_R^{\prime\prime}|\xi_L\sigma_L\sigma_R
\xi_R)\Psi_1(\xi_L\sigma_L\sigma_R\xi_R)=\lambda\Psi_1(\xi_L^{\prime\prime}
\sigma_L^{\prime\prime}\sigma_R^{\prime\prime}\xi_R^{\prime\prime})
\end{equation}
where
\begin{equation}
T(\xi_L^{\prime\prime}\sigma_L^{\prime\prime}\sigma_R^{\prime\prime}
\xi_R^{\prime\prime}|\xi_L\sigma_L\sigma_R\xi_R)=\sum\limits_{\xi_L^{\prime}
\sigma_L^{\prime}\sigma_R^{\prime}\xi_R^{\prime}}T^{(2)}(\xi_L^{\prime\prime}
\sigma_L^{\prime\prime}\sigma_R^{\prime\prime}\xi_R^{\prime\prime}|\xi_L^
{\prime}\sigma_L^{\prime}\sigma_R^{\prime}\xi_R^{\prime})T^{(1)}(\xi_L^
{\prime}\sigma_L^{\prime}\sigma_R^{\prime}\xi_R^{\prime}|\xi_L\sigma_L
\sigma_R\xi_R).
\end{equation}

The eigenvectors at the odd rows $\Psi_2$ follow directly from 
$\Psi_1$
\begin{equation}
\Psi_2= T^{(1)}\Psi_1.
\end{equation}

Optimum size reduction of the matrix $^iT_L\,^iW_B$ is performed by its multiplying
at both sides by rectangular matrices consisting of several eigenvectors of 
a density matrix 
that corresponds to its largest eigenvalues \cite{Whi,Nis}. The density matrix at 
a row $j$ is
constructed from the left and right eigenvectors $\Psi_j,\overline\Psi_j$ \cite{Su3}  
of transfer matrices with their left and right spins, respectively, lying on the
row. For modulated commensurate structures of a period $p$, 
the successive functions $\Psi_j$ 
are not the eigenvectors of one transfer matrix but  of a product of $p$ transfer matrices.
As for ATNNI model we have two different kinds of rows and different
left and right transfer matrices in superblocks,  we need four different density
matrices. The left density matrices have the following forms:
\begin{equation}
\rho_L^{(1)}(\xi_L^a\sigma_L^a|\xi_L^b\sigma_L^b)=\sum\limits_{\sigma_R^c
\xi_R^c}\Psi_1(\xi_L^a\sigma_L^a\sigma_R^c\xi_R^c)\overline\Psi_1(\xi_L^b
\sigma_L^b\sigma_R^c\xi_R^c)
\label{dm1}
\end{equation}
\begin{equation}
\rho_L^{(2)}(\xi_L^{\prime a}\sigma_L^{\prime a}|\xi_L^{\prime b}\sigma_L^
{\prime b})=\sum\limits_{\sigma_R^{\prime c}\xi_R^{\prime c}}\Psi_2(\xi_L^
{\prime a}\sigma_L^{\prime a}\sigma_R^{\prime c}\xi_R^{\prime c})\overline
\Psi_2(\xi_L^{\prime b}\sigma_L^{\prime b}\sigma_R^{\prime c}\xi_R^{\prime c})
\label{dm2}
\end{equation}
In the expressions for the right ones, the summation is performed over the left
spins. 
The functions $\Psi_j$ and $\overline\Psi_j$ are identical that is why the density
matrices in Eqns.\ref{dm1} and \ref{dm2} are symmetric.
Here we should emphasize that the right blocks $T_R$ are not mirror
reflections of the left blocks $T_L$ as they were in the standard approach \cite{Whi}.

By diagonalization of the left symmetric density matrix one obtains a matrix
of orthonormal eigenvectors $O_L$:
\begin{equation}
Q_L(k|\xi\sigma)\rho_L^{(1)}(\xi\sigma|\xi^{\prime}\sigma^{\prime})O_L(\xi^{\prime}
\sigma^{\prime}|\ell)=\omega_k\delta_{k\ell}
\end{equation}
where $Q_L$ is transposed $O_L$ and the eigenvalues $\omega_k$ satisfies
the relation
\begin{equation}
\sum\limits_{k}\omega_k=1.
\end{equation}

Analogously,
we repeat this procedure for the density matrix $\rho_L^{(2)}$ and $\rho_R^{(2)}$
in order to obtain matrices $Q_L^{\prime}$, $O_L^{\prime}$, $Q_R^{\prime}$, and
$O_R^{\prime}$. Discarding half eigenvectors in the matrices $Q$ and $O$ that
correspond to the smallest eigenvalues $\omega_k$, the matrices $O$ and $Q$ can
be used as the reduction matrices in calculation of $^{i+1}T_L^{(1) new}$ via
\begin{equation}
\begin{array}{r}
{}^{i+1}T_L^{(1) new}(\xi_L^{\prime\ new}\sigma_L^{\prime\ new}|\xi_L^{new}\sigma_L^{new})=
\sum\limits_{\xi_L\xi_L^{\prime}\sigma_L\sigma_L^{\prime}}{}^iQ_L^{\prime}(\xi_L^{\prime\ new}|
\xi_L^{\prime}\sigma_L^{\prime}){}^iT_L^{(1)}(\xi_L^{\prime}\sigma_L^{\prime}|\xi_L\sigma_L)\\
{}^iW_B^{(1)}(\sigma_L^{\prime}\sigma_L^{\prime\ new}|\sigma_L\sigma_L^{new})
{}^iO_L(\xi_L\sigma_L|\xi_L^{new})
\label{eq1}
\end{array}
\end{equation}

\begin{figure}[ht]
\centerline{\scalebox{0.9}{\includegraphics{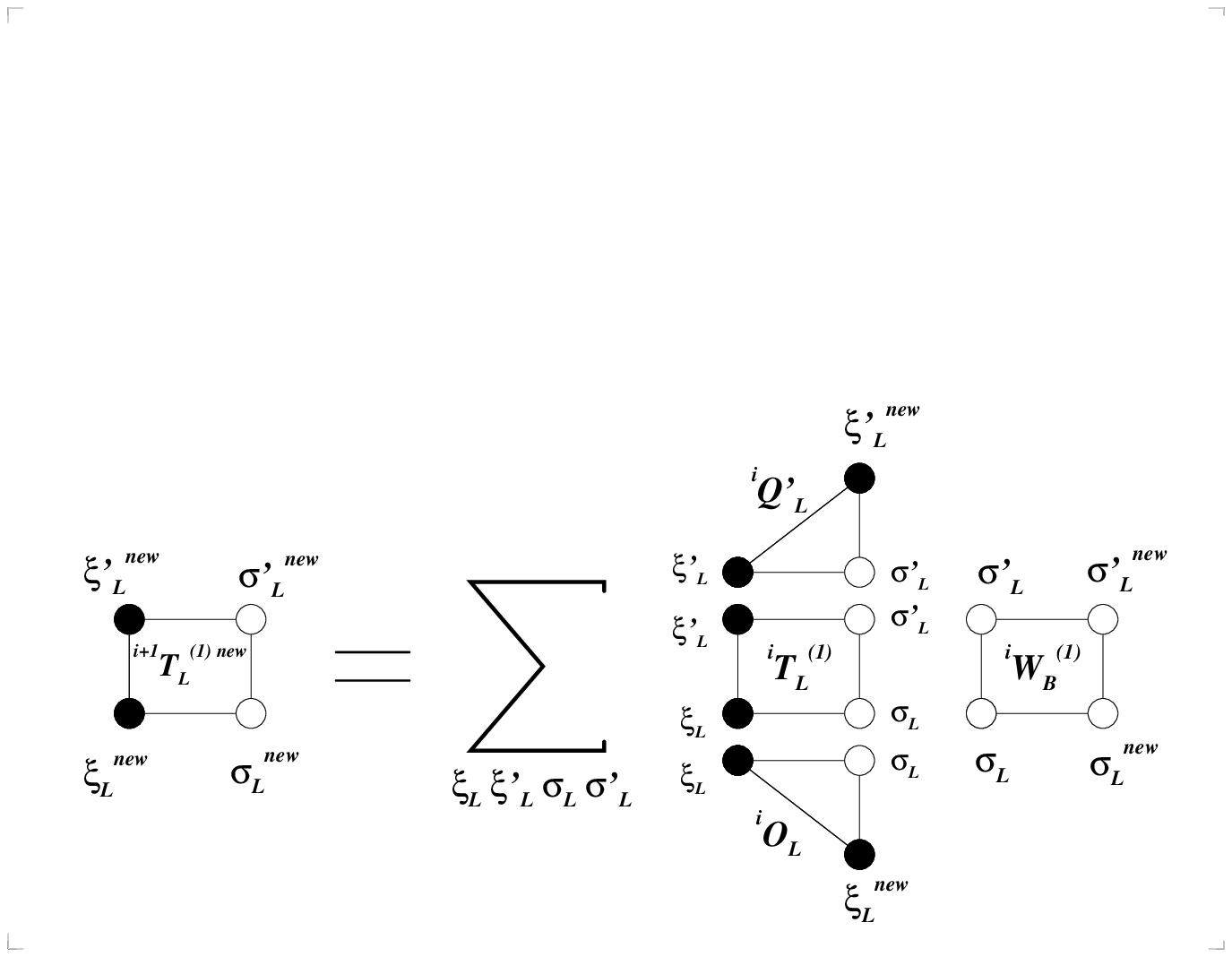}}}
\caption{\it Schematically written equation (\ref{eq1}) which we use for 
computing of the new left
renormalized transfer matrix $T_L^{(1) new}$ from the old one $T_L^{(1)}$.}
\label{blocktransform}
\end{figure}

Generalization for the right block is straightforward. The graphical representation of
Eqn.~\ref{eq1} is in Figure~\ref{blocktransform}.

Notice that at this step (Figure \ref{blocktransform}) the reduction matrix $O_L$ is obtained
from diagonalization of the density matrix $\rho_L^{(1)}$ 
whereas the matrix $Q_L^{\prime}$ 
is taken from the diagonalization of the density matrix $\rho_L^{(2)}$.
From the knowledge of the functions $\Psi_1$, $\Psi_2$, $\overline\Psi_1$, and 
$\overline\Psi_2$ various physical quantities can be found, e.g. site magnetization used in
further calculations
\begin{equation}
<\sigma_L>=\sum\limits_{\xi_L\sigma_L\sigma_R\xi_R}\Psi_1(\xi_L\sigma_L
\sigma_R\xi_R)\sigma_L\overline\Psi_1(\xi_L\sigma_L\sigma_R\xi_R).
\label{spin}
\end{equation}

For the ANNNI model, containing also next-nearest-neighbor interactions, 
the {\it DMRG} method should
be modified further. The left and right block transfer matrices should be indexed, 
besides the block-spin
variable $\xi$, by four spins $\sigma$. The Boltzmann weight is defined 
on a plaquette of at least six spins.
We have constructed the superblock transfer matrix $T$ from the two block transfer 
matrices and the Boltzmann
weight on an eight-site plaquette (Figure \ref{boltzmannannni}) as follows
\begin{equation}
\begin{array}{r}
T(\xi_L\sigma_1\sigma_2\sigma_3\sigma_4\xi_R|\xi_L^{\prime}\sigma_1^{\prime}
\sigma_2^{\prime}\sigma_3^{\prime}\sigma_4^{\prime}\xi_R^{\prime})=
T_L(\xi_L\sigma_1\sigma_2|\xi_L^{\prime}\sigma_1^{\prime}\sigma_2^{\prime})
W_B^{(1)}(\sigma_1\sigma_2\sigma_3|\sigma_1^{\prime}\sigma_2^{\prime}\sigma_3^{\prime})\\
\times
W_B^{(2)}(\sigma_2\sigma_3\sigma_4|\sigma_2^{\prime}\sigma_3^{\prime}\sigma_4^{\prime})
T_R(\sigma_3\sigma_4\xi_R|\sigma_3^{\prime}\sigma_4^{\prime}\xi_R^{\prime})
\end{array}
\end{equation}

\section{Results}

\begin{figure}[!ht]
\centering
\subfigure[\hfill]{\scalebox{0.45}[0.40]{\includegraphics{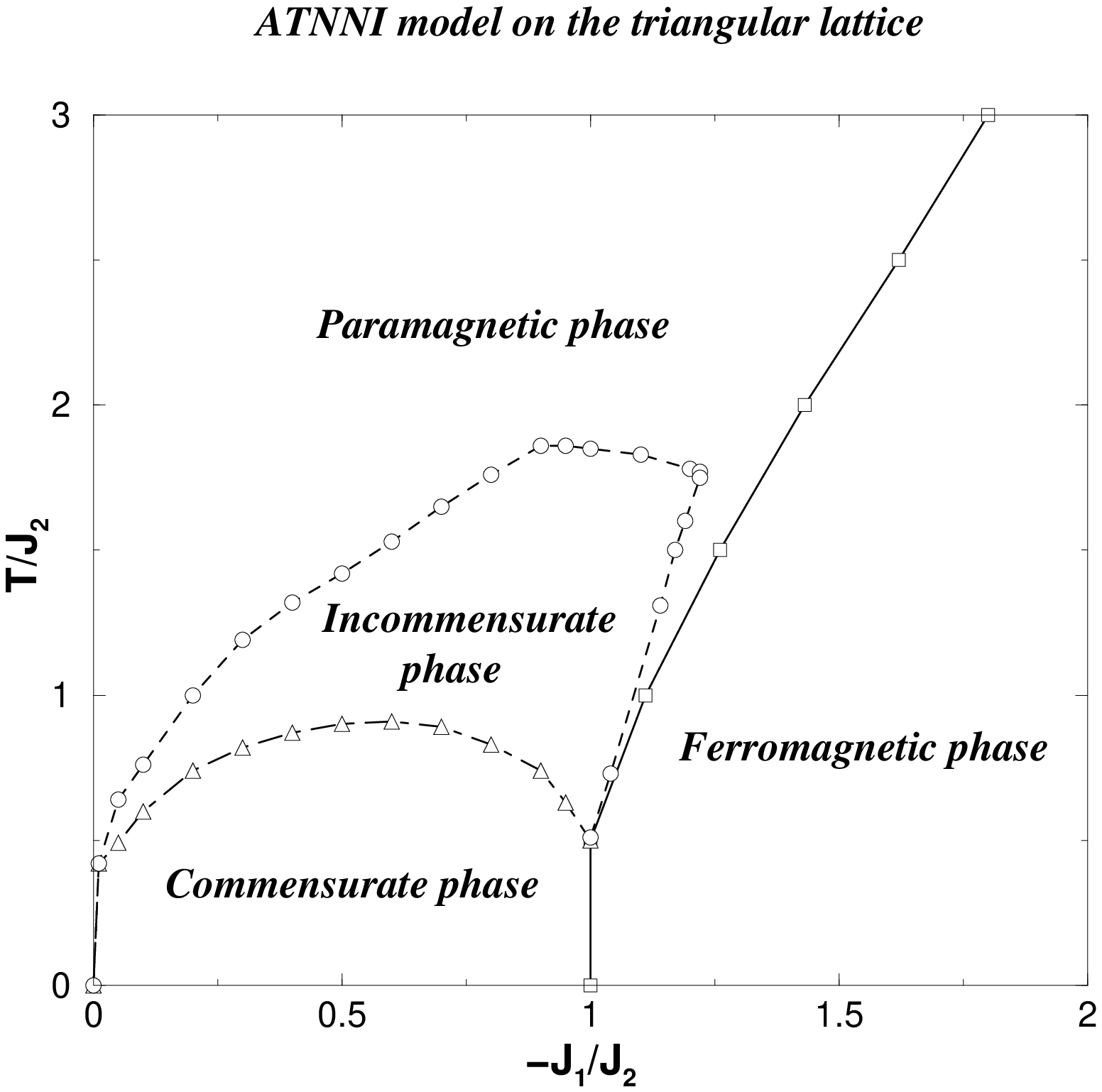}}}
\hfill
\subfigure[\hfill]{\scalebox{0.26}[0.31]{\includegraphics{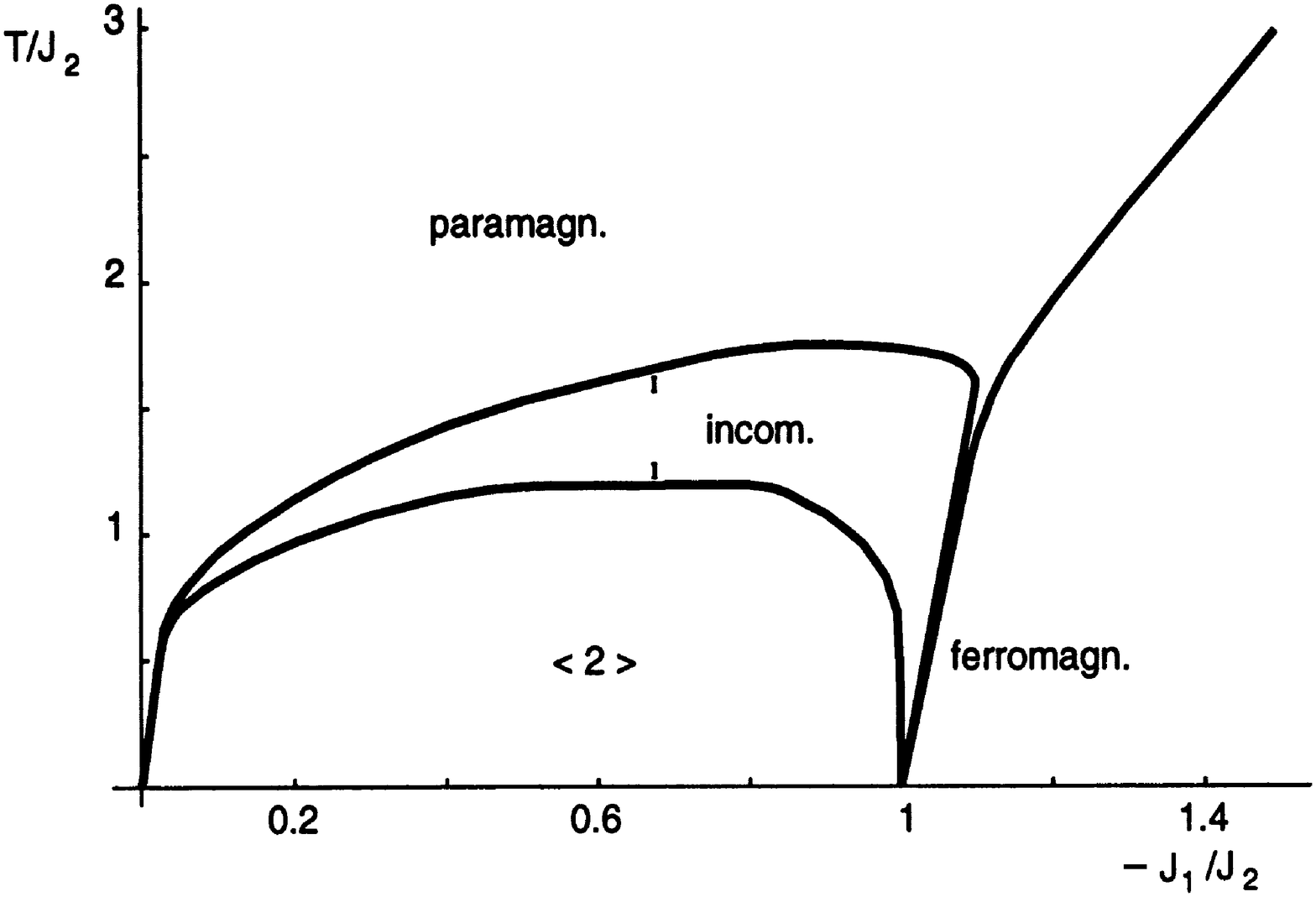}}}
\caption{\it (a) The phase diagram of the ANNNI model obtained by the DMRG method for 
relatively small superblock transfer-matrix size of $N=400$. After increasing the superblock
transfer-matrix size, the IC region becomes narrower and the para-IC phase transition
line is shifted to lower temperatures.
(b) The phase diagram constructed with the cluster transfer matrix method \cite{Sur}.}
\label{annniphasediag}
\end{figure}

\qquad
The properties of the ANNNI model on the triangular lattice were calculated 
recently \cite{Sur} by the 
cluster transfer matrix method \cite{Su2}. The results were consistent with 
numerous calculations
of the ANNNI model on the square lattice. To compare performance of the {\it DMRG} 
method for 
incommensurate (IC) structures with other methods we calculated the phase 
diagram of the ANNNI
model shown in Figure \ref{annniphasediag}(a). The resulting diagram is in 
accordance with previous
calculations (Figure \ref{annniphasediag}(b)). We have confirmed general opinion 
that there is no
Lifshitz point on the ferro-para phase transition line.

The region of the IC structure comes out from the {\it DMRG} rather wide, however, 
we have used a 
low-order approximation ($N=400$). For higher-order approximations the IC--phase region
becomes narrower.                           

\begin{figure}[!ht]
\centerline{\scalebox{0.7}{\includegraphics{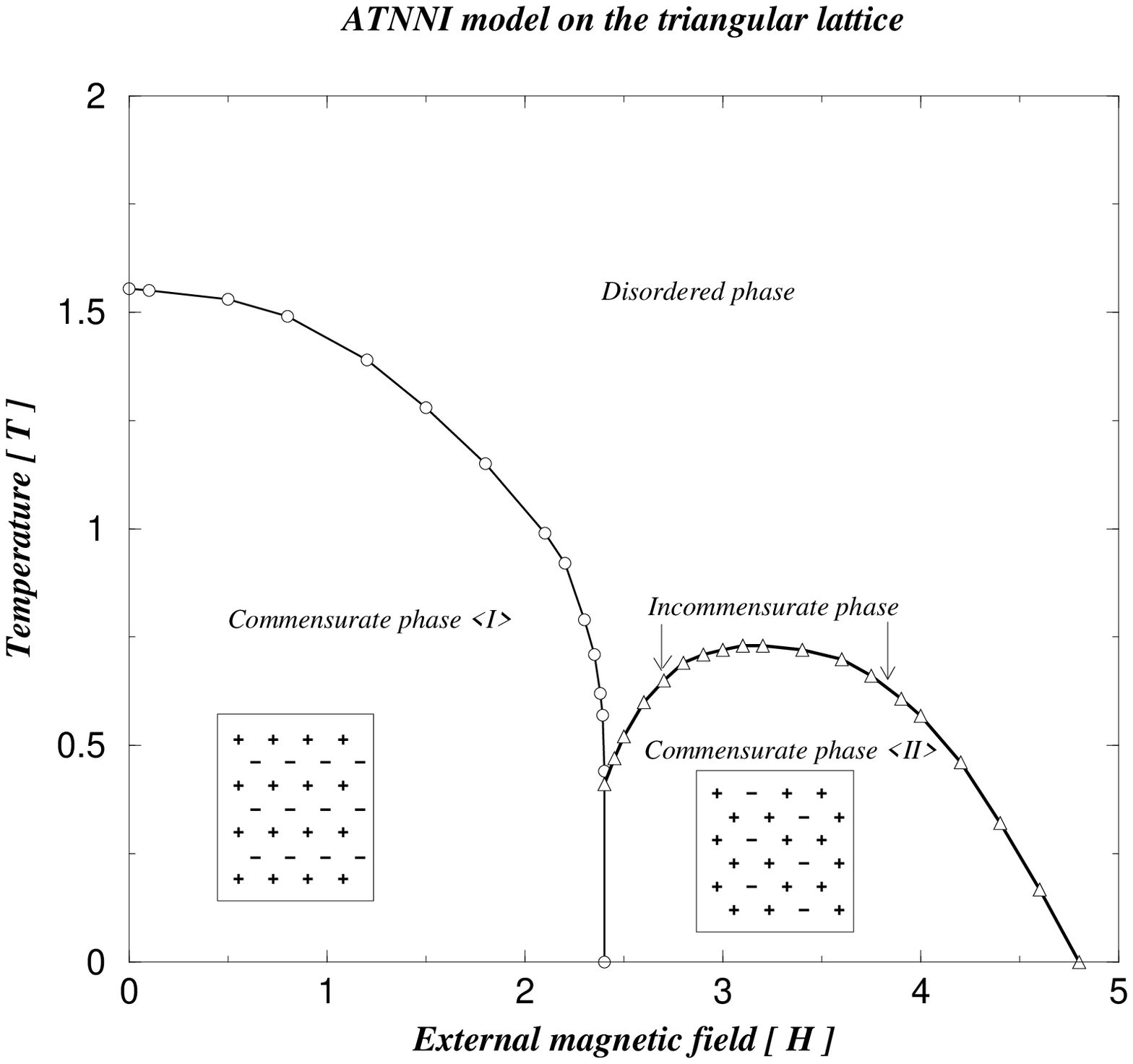}}}
\caption{\it The entire phase diagram of the ATNNI model is constructed by the 
{\it DMRG} technique.
The incommensurate phase appears in narrow region located between the disordered 
phase and
the commensurate phase $\langle II\rangle$ for $2.4<H<4.8$. The ATNNI model is 
highly degenerated for $H=2.4$.}
\label{phaseatnnidiag}
\end{figure}
\begin{figure}[!ht]
\centering
\subfigure[\hfill]{\scalebox{0.6}{\includegraphics{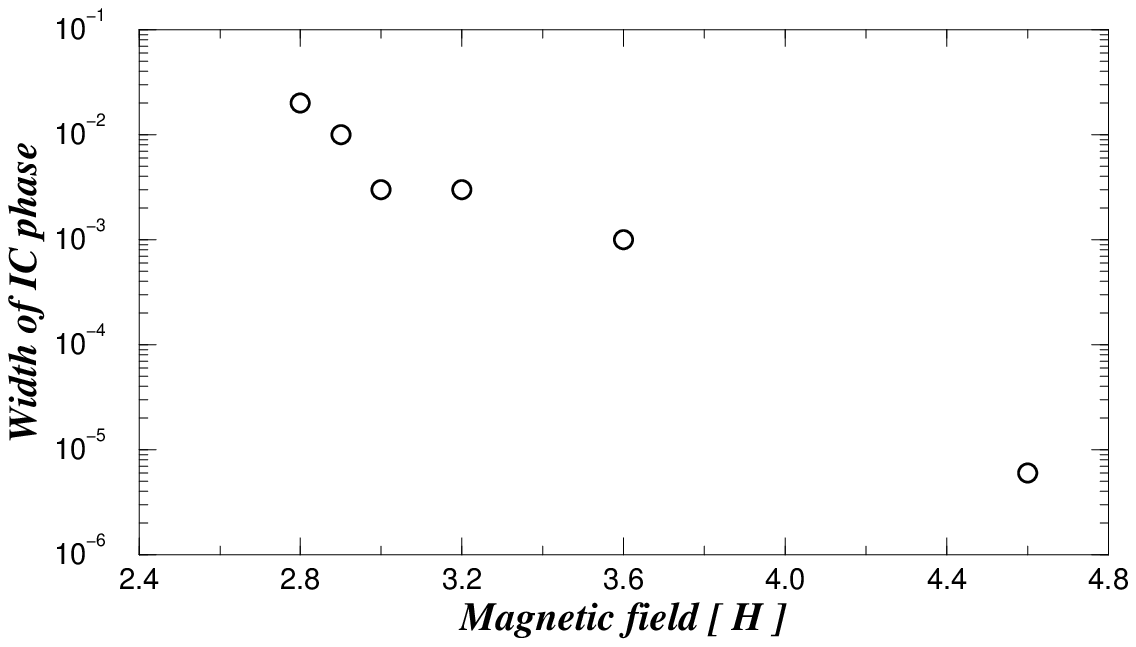}}}
\hfill
\subfigure[\hfill]{\scalebox{0.55}{\includegraphics{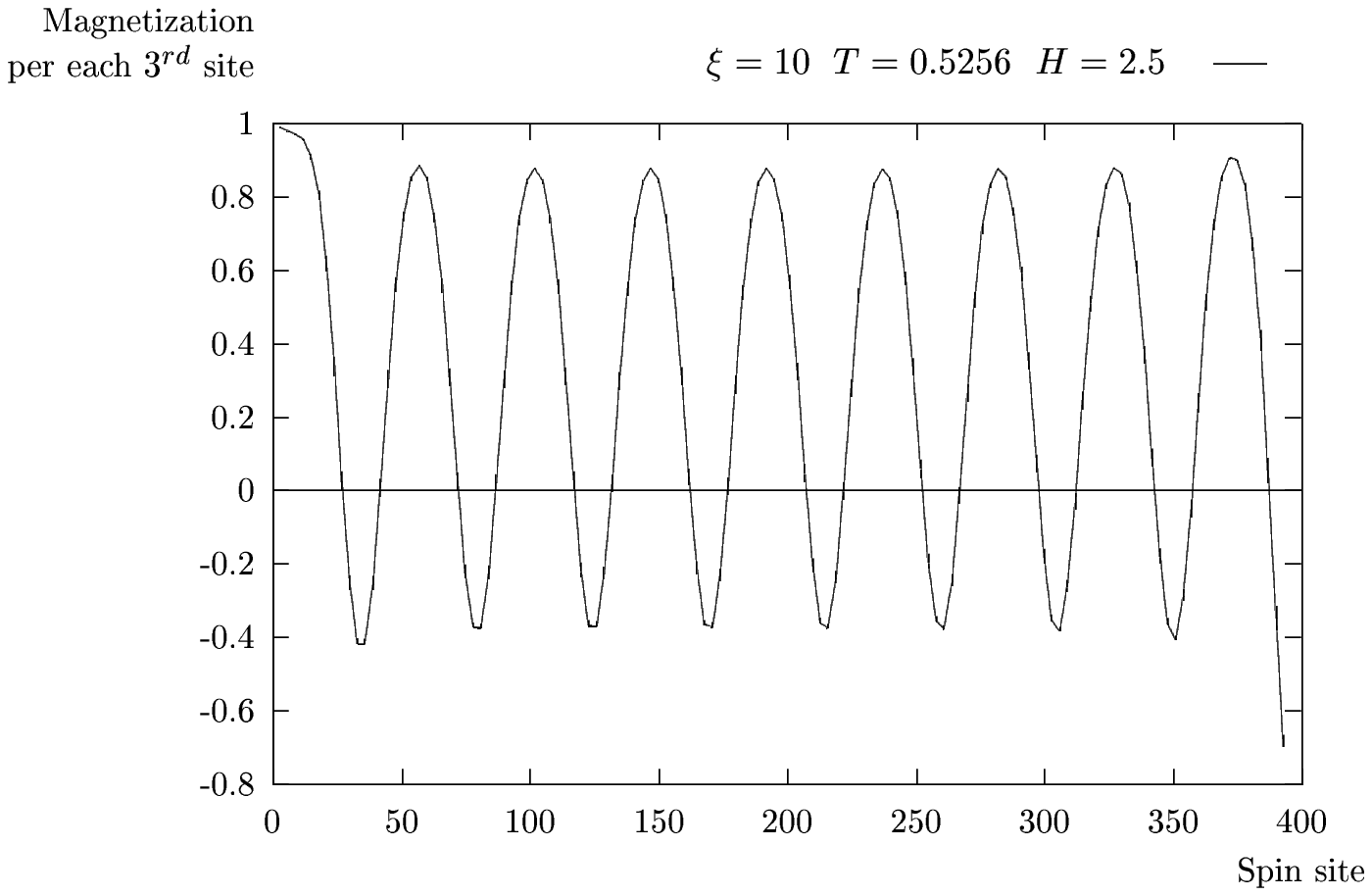}}}
\caption{\it (a) The width of the incommensurate phase measured in units of temperature
vs. external magnetic field $H$. As the field $H$ is increased the distance between the
disordered and commensurate phases $\langle II\rangle$ becomes shorter; (b) Measured
magnetization per each third spin site using Eqn. (\ref{spin}) inside the incommensurate
phase as a function of the spin position on the lattice for $H=2.5$ and $T=0.5256$.}
\label{strip_short}
\end{figure}                 
\begin{figure}[!ht]
\centering
\subfigure[\hfill]{\scalebox{0.4}{\includegraphics{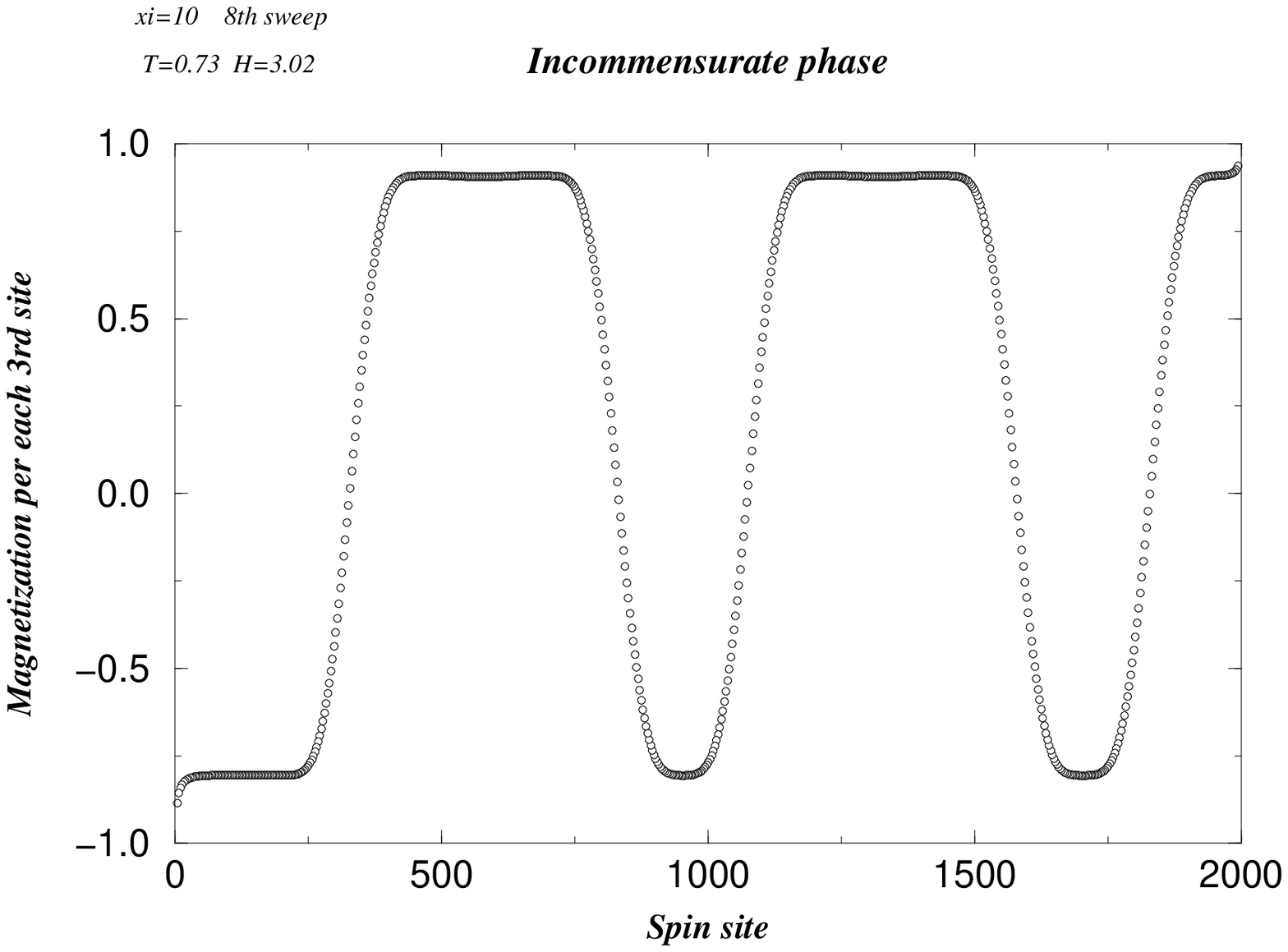}}}
\hfill
\subfigure[\hfill]{\scalebox{0.4}{\includegraphics{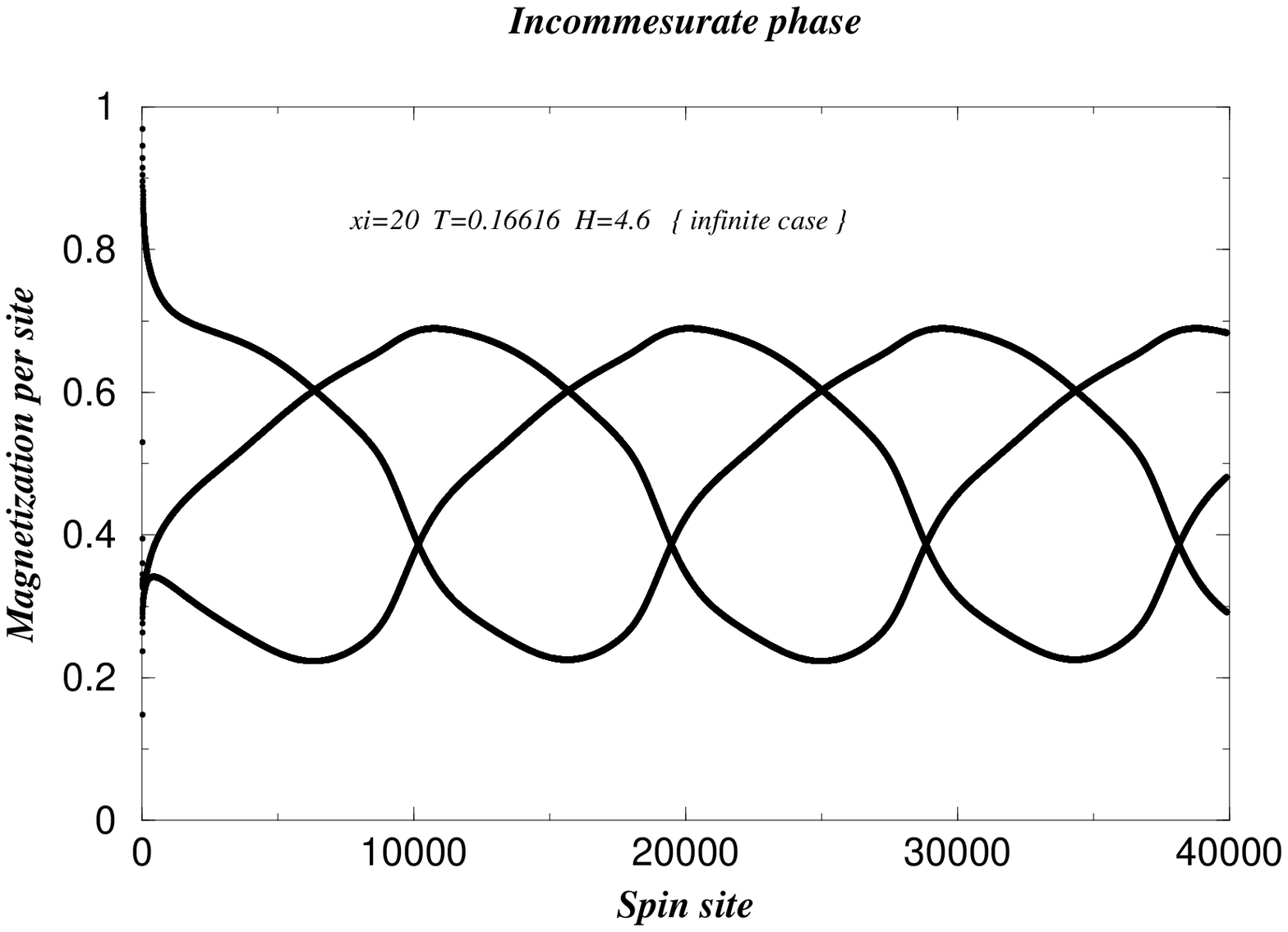}}}
\caption{\it (a) Magnetization vs. position on the lattice measured inside the IC phase 
on each third site for
$H=3.02$ and $T=0.73$; (b)  IC phase obtained  for the large
magnetic field $H=4.6$ was found for temperatures $0.166158<T<0.166162$. The
magnetization is measured on each site.  All three spin waves are plotted.}
\label{medium_long}
\end{figure}

The phase diagram of the ATNNI model (in Figure \ref{phaseatnnidiag}) consists of four
regions (two different commensurate phases, incommensurate and disordered phase). 
Characteristic spin structure
of both commensurate $\langle I\rangle$ and $\langle II\rangle$ phases of the triangular
lattice (Figure \ref{atnnimodel}(a)) is shown in the insets of the same phase diagram.

The phases $\langle I\rangle$ and $\langle II\rangle$ consists of two and three different
sublattices with constant magnetization, respectively. In the IC phase, the magnetization
of each sublattice is periodically modulated and the sublattices become 
equivalent to each other. 
That is why we only plot magnetization of one of these sublattices in some of 
the following Figures
\ref{strip_short} and \ref{medium_long}.

The phase diagram was derived from calculation of  magnetization. The {\it DMRG} method
with spontaneous symmetry breaking yields directly the space modulation of the magnetization
which enables to identify the phase unambiguously. In practical calculations, 
it is even enough
to observe the behavior of the largest eigenvalues of the superblock matrix.  
The period of their spatial modulation is the same as the period of the structure.

The incommensurate structure is floating, i.e. it is not fixed to the 
underlying lattice. In our
calculation with spontaneous symmetry breaking, one of the infinitely many 
positions of the
incommensurate wave is chosen at the beginning of the calculation
and then it remains fixed during the whole further calculation.

We have found the incommensurate structure practically along the whole border 
between the commensurate
 $\langle II\rangle$ phase and the disordered phase. However, in two 
regions the calculations were
inconclusive. The high degeneracy of the ground state at $H=2.4$ and 
$T<0.4$ \cite{Dom}
has also caused highly degenerate largest eigenvalues of the superblock 
transfer matrix, and our 
method did not converge to any periodic structure for magnetic fields 
between 2.40 and 2.41 at 
low temperatures.

The other region, where difficulties were encountered, is located at the 
high-magnetic-field end ($H=4.8$) of the phase diagram. Here, the 
incommensurate phase is 
extremely narrow (Figure \ref{strip_short}(a)) and it has  a very large period (Figure 
\ref{medium_long}(b)). Moreover, due to the proximity of the 
second-order phase transition line,
the convergence is very slow.

We have started the calculation with the {\it ISM} where the superblock 
transfer matrix is constructed 
from left and right transfer matrices of the {\it previous iteration step}. 
After a large enough number
of iterations  we obtained a final result for commensurate structures 
including the disordered
phase. For the incommensurate structure it is necessary to perform afterwards 
some sweeps of the
{\it FSM}, where for the left sweep, the left transfer matrices are 
taken from the {\it previous sweep}
(analogously, for the right sweep).

The IC structure appears already after the application of the {\it ISM} 
but the correct shape of the
magnetization is acquired after the {\it FSM}, only.

The shape of period of the IC structure changes with magnetic field $H$ and temperature $T$.
The period of the IC structure increases with the increasing magnetic field 
and decreasing temperature.
At low temperatures (close to $\langle II\rangle$--IC phase transition line)
 the structure consists of 
wide domains of the phase $\langle II\rangle$ separated by narrow domain walls. 
At higher temperatures
near to disorder--IC transition, the domain walls become wider, the period shorter 
and the structure
acquires a sinusoidal-like shape. 

Both phase transitions are continuous.
Inverse period of the structure and wave amplitude tend to zero 
at the $\langle II\rangle$--IC and 
didorder--IC phase transition lines, respectively. It should be noted that the notions of 
low and high
temperatures must be understood within an extremely narrow temperature interval 
where the IC phase
exists.

The effect of magnetic field on the IC phase is similar (but inverse) to the temperature
effects. 
Low magnetic
field (near $2.4$) enhances the high temperature effects,
 while the high magnetic field
(near $4.8$) the low temperature ones.

For the magnetic field $H$ close to the value of $4.8$, the period is very 
long, that is why we were 
able to perform the {\it ISM} only with an incorrect magnetization shape 
which would need  a further improvement 
with the {\it FSM} (Figure \ref{medium_long}(b)).

Our calculations converged to the stable periodic solution at the most part 
of the commensurate
$\langle II\rangle$-disordered phase borders. Hers, the IC phase has been found everywhere. 
This fact
leads us to a conjecture (in contrary to \cite{Dom}) that the Lifshitz point 
does not exists in the ATNNI
model.

\section{Summary}

\qquad
The {\it DMRG} method has been used to investigate incommensurate structures 
in 2D classical
model for the first time. We found that it reproduces well the previous results 
for the ANNNI model.
In the case of the  ATNNI model it has shown much better performance in the regions 
where the previous
approaches (the cluster transfer matrix method \cite{Su2,Tot} near $H=2.4$ and 
the free-fermion
approximation \cite{Dom} for $H>3$) has failed.

In \cite{Dom} on the basis of scaling properties of Monte Carlo calculations and the exact
diagonalization of finite strips, the authors concluded that at $H\cong 3$ the IC 
structure disappears
and at higher fields $H$ the direct phase transition 
between commensurate $\langle II\rangle$ and 
disordered phases is continuous.

We have observed the IC phase everywhere between the disordered and 
commensurate $\langle II\rangle$
phases, i.e. we have found no Lifshitz point where the three phases meets: commensurate, 
incommensurate, and disordered. Nevertheless, measured widths of the IC phase are extremely
small at large $H$ and exponentially tend to zero at $H=4.8$. As the width of the IC phase 
gets
narrower for the high-order approximations we cannot completely exclude the scenario of
Domany and Schaub \cite{Dom}.

Our belief in correct description of incommensurate phases by the {\it DMRG} technique 
is supported
by the reproduction of the ANNNI phase diagram with generally expected features.

\section*{Acknowledgments}

\qquad
This work has been supported by Slovak Grant Agency, Grant n. 2/4109/98.
We would like to thank the organizers of the DMRG Seminar and Workshop in Dresden for 
the opportunity to participate in the meetings, especially for the useful discussion with
T.~Nishino and I.~Peschel. A.~G. also thanks P.~Marko\v{s} for useful discussions and 
comments.



\begin{thebibliography}{99}
\bibitem{Whi} S.~R.~White, Phys. Rev. Lett. {\bf 69}, 2863 (1992); Phys. Rev. 
B {\bf 48} 10345 (1993).
\bibitem{Nis} T.~Nishino, J. Phys. Soc. Jpn. {\bf 64}, 3598 (1995).
\bibitem{Var} S.~\"Ostlund and S.~Rommer, Phys. Rev. Lett. {\bf 75}, 3537 (1995);
Phys. Rev. B {\bf 55}, 2164 (1997); K.~Okunishi, Y.~Hieida and Y.~Akutsu, 
cond--mat/9810239.
\bibitem{Sur} P.~Pajersk\' y and A.~\v{S}urda, J. Phys. A: Math. Gen. {\bf 30} 4187
(1997).
\bibitem{Su3} A.~\v{S}urda, Acta Phys. Slov. {\bf 49}, 325 (1999).
\bibitem{Kar} K.~Hallberg, cond--mat/9910082.
\bibitem{Drw} E.~Carlon and A.~Drzewi\`nski, Phys. Rev. Lett. {\bf 79}, 1591 (1997).
\bibitem{Ni2} T.~Nishino and K.~Okunishi, J. Phys. Soc. Jpn. {\bf 64} 4084 (1995).
\bibitem{CTM} R.~Baxter, J. Math. Phys. {\bf 9}, 650 (1968); J. Stat. Phys. {\bf 19},
461 (1978).
\bibitem{CTMRG} T.~Nishino and K.~Okunishi, J. Phys. Soc. Jpn. {\bf 65} 891 (1996); 
Phys. Soc. Jpn. {\bf 66} 3040 (1997); T.~Nishino, K.~Okunishi and M.~Kikuchi, Phys.
Lett. A {\bf 213}, 69 (1996).
\bibitem{3DN} T.~Nishino and K.~Okunishi, J. Phys. Soc. Jpn. {\bf 67}, 3066 (1998).
\bibitem{Dom} E.~Domany and B.~Schaub, Phys. Rev. B {\bf 29} 4095 (1983).
\bibitem{Vil} J.~Villain and P.~Bak, J. Phys. {\bf 42}, 657 (1981).
\bibitem{Pok} V.~L.~Pokrovsky and A.~L.~Talapov, Phys. Rev. Lett. {\bf 42}, 65 (1979).
\bibitem{Sch} H.~J.~Schultz, Phys. Rev. B {\bf 22}, 5274 (1980).
\bibitem{Bar} M.~N.~Barber and P.~M.~Duxbury, J. Phys. A {\bf 14}, L251 (1981).
\bibitem{Fis} M.~E.~Fischer and W.~Selke, Phys. Rev. Lett. {\bf 44}, 1502 (1980).
\bibitem{Hal} F.~D.~M.~Haldane, P.~Bak, and T.~Bohr, Phys. Rev. B {\bf 28}, 2743 (1983).
\bibitem{And} A.~Gendiar and A.~\v{S}urda in preparation.
\bibitem{Car} E.~Carlon, M.~Henkel, and U.~Schollw\"{o}ck, Europhys. J. B to appear (1999).
\bibitem{Nig} P.~Nightingale, J. Appl. Phys. {\bf 53}, 7927 (1982).
\bibitem{Bax} R.~J.~Baxter, Exactly Solved Models in Statistical Physics, 
Academic Press, London (1982).
\bibitem{Sel} W.~Selke, Phys. Rep. {\bf 170}, 213 (1988); W.~Selke, {\it Phase Transition and 
Critical Phenomena} vol. {\bf 15} (New York: Academic) (1992).
\bibitem{Su2} I.~Karasov\'a and A.~\v{S}urda, J. Stat. Phys. {\bf 70}, 675 (1993).
\bibitem{Tot} L.~T\'{o}th and A.~\v{S}urda (unpublished).

\end{thebibliography}
\end{document}